\documentclass[aps,prd,superscriptaddress,twocolumn]{revtex4-1}
\usepackage[utf8]{inputenc}
\usepackage{dsfont}
\usepackage{amsfonts}
\usepackage{amsmath}
\allowdisplaybreaks[4]        
\usepackage{amssymb}
\usepackage{euscript}     
\usepackage{braket}
\usepackage{starfont}
\usepackage{color,soul}         
\usepackage{tensor}        
\usepackage{amsthm}
\usepackage{graphicx}
\usepackage{slashed}
\usepackage{leftidx}
\usepackage{subfigure}
\usepackage{bbm}
\definecolor{outerspace}{rgb}{0.25, 0.29, 0.3}
\definecolor{scarlet}{rgb}{1.0, 0.13, 0.0}
\usepackage[header,title,page,titletoc]{appendix}  
\definecolor{princetonorange}{rgb}{1.0, 0.56, 0.0}
\definecolor{WildStrawberry}{rgb}{1.0, 0.26, 0.64}
\definecolor{rossocorsa}{rgb}{0.83, 0.0, 0.0}
\definecolor{navyblue}{rgb}{0.0, 0.0, 0.5}
\usepackage{float}
\usepackage[paper=letterpaper,margin=1in]{geometry}
\usepackage{mathtools}

\DeclareMathAlphabet{\pazocal}{OMS}{zplm}{m}{n}
\newcommand{\req}[1]{(\ref{#1})} 

\newcommand{\bea}{\begin{eqnarray}}
\newcommand{\diff}{\mathrm{d}}
\newcommand{\eea}{\end{eqnarray}}
\newcommand{\ba}{\begin{eqnarray}}
\newcommand{\ea}{\end{eqnarray}}

\newcommand{\be}{\begin{equation}}
\newcommand{\ee}{\end{equation} }
\newcommand{\beqa}{\begin{eqnarray}}
\newcommand{\eeqa}{\end{eqnarray}}
\newcommand{\beqar}{\begin{eqnarray*}}
\newcommand{\eeqar}{\end{eqnarray*}}

\renewcommand{\req}[1]{(\ref{#1})}

\newcommand{\ie}{{\it i.e.,}\ }

\makeatletter
\newcommand{\dal}{\mathop{\mathpalette\dal@\relax}}
\newcommand{\dal@}[2]{%
  \begingroup
  \sbox\z@{$\m@th#1\square$}%
  \dimen0=\fontdimen8
    \ifx#1\displaystyle\textfont\else
    \ifx#1\textstyle\textfont\else
    \ifx#1\scriptstyle\scriptfont\else
    \scriptscriptfont\fi\fi\fi3
  \makebox[\wd\z@]{%
    \hbox to \ht\z@{%
      \vrule width \dimen0
      \kern-\dimen0
      \vbox to \ht\z@{
        \hrule height \dimen0 width \ht\z@
        \vss
        \hrule height 2\dimen0
      }%
      \kern-2.5\dimen0
      \vrule width 2.5\dimen0
    }%
  }%
  \endgroup
}
\makeatother

\usepackage{hyperref}
\hypersetup{
    colorlinks,
    citecolor=navyblue,
    filecolor=navyblue,
    linkcolor=navyblue,
    urlcolor=navyblue
}

\begin{document}

\title{Nonlocal Massive Gravity from Einstein Gravity}
\author{Pablo Bueno}
\email{pablobueno@ub.edu}
\affiliation{Departament de F\'isica Qu\`antica i Astrof\'isica, Institut de Ci\`encies del Cosmos\\
 Universitat de Barcelona, Mart\'i i Franqu\`es 1, E-08028 Barcelona, Spain }

\author{Pablo A. Cano}
\email{pablo.cano@icc.ub.edu}
\affiliation{Departament de F\'isica Qu\`antica i Astrof\'isica, Institut de Ci\`encies del Cosmos\\
 Universitat de Barcelona, Mart\'i i Franqu\`es 1, E-08028 Barcelona, Spain }

\author{Robie A. Hennigar}
\email{robie.hennigar@icc.ub.edu}
\affiliation{Departament de F\'isica Qu\`antica i Astrof\'isica, Institut de Ci\`encies del Cosmos\\
 Universitat de Barcelona, Mart\'i i Franqu\`es 1, E-08028 Barcelona, Spain }


\begin{abstract}
We present a top-down construction of a three-dimensional non-local theory of massive gravity. This “Non-Local Massive Gravity” (NLMG) is obtained as the gravitational theory induced by Einstein gravity on a brane inserted in Anti-de Sitter space modified by an overall minus sign. The theory involves an infinite series of increasingly complicated higher-derivative corrections to the Einstein-Hilbert action, with the quadratic term coinciding with New Massive Gravity. We obtain an analytic formula for the quadratic action of NLMG and show that its linearized spectrum consists of an infinite tower of positive-energy massive spin-2 modes. We compute the Newtonian potential and show that the introduction of the infinite series of terms makes it behave as $\sim 1/r$ at short distances, as opposed to the logarithmic behavior encountered when the series is truncated at any finite order. We use this and input from brane-world holography to argue that the theory may contain asymptotically flat black hole solutions.

\end{abstract}
\maketitle

Gravity is simpler in three dimensions. The Riemann tensor is fully determined by the Ricci tensor and all solutions of Einstein gravity are  
 locally maximally symmetric. Additionally, local dynamics is trivial in the absence of matter, as metric perturbations propagate no degrees of freedom \cite{Deser:1983tn}.
 
Non-trivial dynamics can be obtained by adding higher-curvature corrections to the Einstein-Hilbert action \cite{Sisman:2011gz,Gullu:2010vw}. The prototypical instance corresponds to ``New Massive Gravity'' (NMG) \cite{Bergshoeff:2009hq,Bergshoeff:2009fj}, where a smart choice of quadratic correction gives rise to a theory that propagates a  unitary massive graviton. The construction relies on an unusual overall minus sign in the action which heals the otherwise ghost-like mode associated to the quadratic term. This procedure would transform the Einstein graviton into a  ghost in higher dimensions, but not in three, where it does not exist at all.

In $D=4$, quadratic corrections render Einstein gravity perturbatively renormalizable  at the price of making it non-unitary \cite{Stelle:1976gc}. On the other hand, NMG is power-counting UV-finite and unitary, although ultimately non-renormalizable  perturbatively  \cite{Deser:2009hb, Muneyuki:2012ur}.

Numerous generalizations of NMG have been studied in the literature attending to different criteria. 
One direction entails demanding the corresponding theories to admit a holographic c-theorem \cite{Sinha:2010ai,Paulos:2010ke,Paulos:2012xe,Bueno:2022lhf}. This includes a Born-Infeld-type extension of NMG \cite{Gullu:2010pc}. Additional theories are selected by the condition that they can be written in a Chern-Simons-like form \cite{Bergshoeff:2014bia,Afshar:2014ffa,Bergshoeff:2021tbz}.  Yet a different route involves considering certain $D\rightarrow 3$ limits of higher-dimensional theories \cite{Alkac:2020zhg,Alkac:2022zda}. Alternative ideas include 
\cite{Banados:2009it,Bergshoeff:2013xma,Bergshoeff:2014pca,Alkac:2017vgg,Alkac:2018eck,Ozkan:2018cxj,Afshar:2019npk}. Despite their variety, all these approaches follow from bottom-up considerations.

In this paper we present a unitary, non-local extension of NMG which we dub ``Non-Local Massive Gravity'' (NLMG). Our construction is top-down in the sense that the theory is uniquely defined as the gravitational action induced on a codimension-1 brane inserted near the boundary of Anti-de Sitter (AdS) spacetime by four-dimensional Einstein gravity, modified by an overall minus sign, namely,
\begin{equation}\label{nlmg1}
I_{\rm NLMG}\equiv -I_{\rm bgrav}\, .
\end{equation}
The above action involves an infinite tower of higher-curvature corrections to the Einstein-Hilbert action, 
\begin{align}\label{totalactionn}
I_{\rm NLMG}=&-\int \frac{\diff^{3}x\sqrt{-g}}{16\pi G_3 } \left[ R+ \sum_{n=2}^{\infty} \ell^{2n-2} \mathcal{L}_{(n)}\right]\, ,
\end{align}
where $\ell$ is a gravitational coupling with dimensions of length. The quadratic and cubic densities read, respectively,
\begin{align}
\mathcal{L}_{(2)}&=-R_{ab}R^{ab} +\frac{3}{8}R^2\, , \\ \notag
\mathcal{L}_{(3)}&=-\frac{2}{3}\left[4R_a^bR_b^cR_c^a-\frac{29}{8}R R_{ab}R^{ab}+\frac{49}{64}R^3 \right. \\  &\qquad\quad\,  \left. -R_{ab} \Box R^{ab}+ \frac{3}{8} R \Box R\right]\, .
\end{align}
$\mathcal{L}_{(2)}$ is nothing but the NMG density, whereas higher orders give rise to an increasingly complicated structure of densities which involve the Ricci tensor and its covariant derivatives \cite{Kraus:1999di}. Explicit expressions up to $n=6$ can be found in \cite{Bueno:2022log}, where it has been conjectured that every density can always be written as $\mathcal{L}_{(n)}=\mathcal{S}_n[R_{ab}]+T_n[\nabla_a,R_{ab}]$, where the piece involving covariant derivatives vanishes on conformally flat backgrounds and $\mathcal{S}_n[R_{ab}]$ has second-order equations on cosmological/domain-wall ansatze  --- see also \cite{Anastasiou:2020zwc,Moreno:2023arp}.\footnote{There are some additional features which these densities satisfy. $\mathcal{L}_{(2)}$ is the only three-dimensional higher-curvature density such that the trace of its equations of motion is second-order in derivatives \cite{Oliva:2010zd}. Additionally, $\mathcal{S}_3[R_{ab}]$ had been previously identified as the cubic generalization of NMG which admits an holographic c-theorem \cite{Paulos:2010ke}. Finally, $T_3[\nabla_a,R_{ab}]$ is proportional to the Cotton tensor squared \cite{Bueno:2022log}, which is the only three-dimensional higher-curvature density such that the trace of its equations of motion is third-order in derivatives \cite{Oliva:2010zd}.}

As we show below, the linearized spectrum of NLMG consists of an infinite tower of massive spin-2 modes that carry positive energy. This follows from an analysis of the quadratic action of the theory, which can be written in closed-form in terms of elementary functions of the Laplace operator --- see \req{totalactionn}. We find an analytic formula for the Newtonian potential and show that the infinite tower of terms has the effect of producing a $\sim 1/r$ behavior near the point-like source. This remarkably differs from the logarithmic behavior encountered when the series is truncated at any finite order and suggests the existence of a new class of three-dimensional asymptotically-flat black-hole solutions.  

{\bf NLMG from brane-world gravity:}
The origin of the higher-curvature densities appearing in the action of NLMG can be understood as follows.
Consider Einstein gravity in the presence of a negative cosmological constant in four dimensions. Inserting a brane near the AdS boundary allows to reinterpret the bulk theory in terms of an induced gravitational theory on the brane coupled to a cutoff CFT \cite{Verlinde:1999fy, Gubser:1999vj, deHaro:2000wj}.\footnote{see also~\cite{Geng:2023qwm} for recent developments on this subject.} Schematically, one has
$
I_{\rm EH}+I_{\rm brane}=I_{\rm bgrav}+I_{\rm CFT}\, .
$
In this expression, $I_{\rm EH}$ is the Einstein-Hilbert action (plus the Gibbons-Hawking boundary term), $I_{\rm brane}$ is the brane world-volume action, $I_{\rm CFT}$ is the cutoff CFT, and $I_{\rm bgrav}$ is the brane-world gravity action previously mentioned.

The brane-world theory is defined by projecting the bulk Einstein equations on the brane and then decomposing them on intrinsic curvature and extrinsic curvature --- which plays the role of the brane stress-energy tensor. One of the components of the bulk Einstein equations yields\footnote{Here we are setting the brane tension to a particular value such that the resulting brane-world theory has zero cosmological constant. However, AdS or dS brane-worlds are possible if one chooses the tension differently.}
\begin{equation}\label{isra}
\Pi =\frac{\ell }{2}\left[R+\Pi_{ab}\Pi^{ab}-\frac{1}{2}\Pi^2 \right]\, ,
\end{equation}
where $R$ is the Ricci scalar of the induced metric on the brane, $\Pi^{ab}\equiv \frac{2}{\sqrt{-g}}\frac{\delta}{\delta g^{ab}} I_{\rm bgrav}$
 is the equation of motion of the brane-world theory\footnote{We remark that $\Pi^{ab}$ appears here as because it its relation to the extrinsic curvature \cite{PhysRevD.47.1407}.}, $\Pi$ its trace and $\ell$ is the AdS radius in the bulk. As shown by \cite{Kraus:1999di}, the relation \req{isra} is enough to derive the form of the induced gravity action $I_{\rm bgrav}$.
 Writing the brane-world Lagrangian as a series expansion in derivatives of the metric, yields a recursive relation for the trace that has the form

\begin{align} \label{recu}
\Pi_{(n\geq 2)}&=\frac{1}{2}\sum_{i=1}^{n-1}\left[\Pi_{(i)\,ab}\Pi^{ab}_{(n-i)}-\frac{\Pi_{(i)} \Pi_{(n-i)}}{2} \right]\, ,
\end{align}
where  $\Pi^{ab}_{(n)}$ is the equation of motion of the $n$-th density. 
The above relation is complemented with the seed condition, $\Pi_{(1)}=R/2$,  together with 
\begin{equation}\label{pin}
\Pi_{(n)}=(3-2n)\mathcal{L}_{(n)} \, , 
\end{equation}
which holds up to total derivatives and follows from consistency of the induced gravity action under Weyl rescalings of the metric \cite{Kraus:1999di}. These expressions allow one to unambiguously  determine, order by order, the full tower of Lagrangian densities --- see \cite{Kraus:1999di, Bueno:2022log}. 

Observe that the Einstein-Hilbert piece of the NLMG action contains a somewhat unusual negative sign which, just like in the NMG case \cite{Bergshoeff:2009hq}, is required for the linearized massive modes of the theory to have positive kinetic energy --- see below. While this would make the usual massless graviton become a ghost in higher dimensions, in $D=3$ no such mode exists in the linearized spectrum --- it is pure gauge --- which saves the consistency of the construction \cite{Deser:1983tn}.

Note also that even though $I_{\rm NLMG}$ is an extremely complicated gravitational Lagrangian, the special relations satisfied by the equations of motion can give us some hints on the solutions of the theory. In particular, observe that
\begin{equation}
\Pi^{ab}=0\quad  \Longrightarrow\quad  R=0\, ,
\end{equation}
namely, all solutions of NLMG have a vanishing Ricci scalar, a property which is obviously shared by Einstein gravity, but not by NMG.

{\bf Quadratic action:} The linearization of the theory around flat space is useful for constructing perturbative solutions and is essential for assessing the spectrum of the theory. The linearization of a higher-curvature theory can always be reduced to the problem of obtaining an equivalent \textit{quadratic action} which involves all terms with two powers of curvature tensors~\cite{Hindawi:1995an, Gullu:2010em}. In the case of NLMG, the quadratic action can be straightforwardly obtained from the results presented in \cite{Aguilar-Gutierrez:2023kfn}. It can be written in the remarkably compact form
\begin{align}\label{totalactionn}
I_{\rm NLMG}^{(2)}=&-\int \frac{ \diff^{3}x\sqrt{-g}}{16\pi G }  \bigg[R\\ \notag &-\ell^2 R^{ab}F \left(\ell^2\Box\right) \left( R_{ab}-\frac{3}{8} g_{ab }R\right)\bigg]\, ,
\end{align}
where we defined
\begin{equation}\label{Fxx}
F(x)\equiv\frac{\sin \left(\sqrt{x}\right)}{x \sin \left(\sqrt{x}\right)+\sqrt{x} \cos \left(\sqrt{x}\right)} \, .
\end{equation}
and $\Box=g^{ab}\nabla_{a}\nabla_{b}$. 
Similarly to the full NLMG action, its quadratic version also provides a non-local completion of  NMG. Indeed, expanding around $x=0$, one finds
$F(x) \approx 1 -\frac{2}{3} x + \frac{7}{15} x^2+\dots $, so the first term in the infinite expansion yields the usual NMG density \cite{Bergshoeff:2009hq}.

{\bf Linearized spectrum:}
Let us now study the linearized equations of NLMG on a Minkowski background --- which is itself a trivial solution of the equations of motion. Hence, we consider an expansion of the metric of the form 
\begin{equation}
g_{ab}=\bar g_{ab} +h_{ab}
\end{equation}
where $\bar g_{ab}$ is the Minkowski metric and $h_{ab}$ is a small perturbation. Using the method developed in \cite{Bueno:2016ypa}, it is straightforward to obtain the linearized equations of the NLMG theory. The result reads
\begin{align}\label{tens}
\frac{1}{32\pi G} \left[-1 +F(\ell^2 \bar \Box)\ell^2\bar\Box\right]G_{ab}^{(1)} &=0\, , \\ \label{sca}
R^{(1)}&=0\, ,
\end{align}
where $G_{ab}^{(1)}$  and $R^{(1)}$ are the linearized  Einstein tensor and Ricci scalar, respectively,
\begin{align}\notag
G_{ab}^{(1)}&=\bar\nabla_{(a|}\bar\nabla^ch_{c|b)}-\frac{1}{2}\bar\Box h_{ab}-\frac{1}{2}\bar\nabla_a\bar\nabla_b h-\frac{1}{2}\bar g_{ab}R^{(1)} ,\\
R^{(1)}&=\bar\nabla^a\bar\nabla^bh_{ab}-\bar\Box h\, .
\end{align}
Eq.\,\req{sca} implies that the linearized metric propagates no scalar modes, a feature shared by brane-world gravities in general dimensions as well as by the usual NMG theory.

Fixing the harmonic gauge, $\bar \nabla h_{ab}=\frac{1}{2} \bar \nabla_b h$, the vanishing of the linearized Ricci scalar can be used to set $h\equiv \bar g^{ab}h_{ab}=0$ and the linearized Einstein tensor simply becomes $G_{ab}^{(1)}=\frac{1}{2}\bar\Box h_{ab}$. Inserting this in \req{tens} and Fourier-transforming  the resulting expression we can read off the propagator of the theory. Up to a fixed tensorial structure \cite{Muneyuki:2012ur}, the result is
\begin{equation}
P_{\rm NLMG}(k)=-\frac{1}{\ell^2 k^2}+\frac{{\rm tanh}(\ell k)}{\ell k}\, .
\end{equation}
For each pole we have $k^2=-m^2$, where $m^2$ is the squared mass of the corresponding mode. There is an obvious pole at $k^2=0$, corresponding to the would-be massless graviton of Einstein gravity, which is pure gauge in three dimensions. Additionally, one finds an infinite tower of poles corresponding to  massive spin-2 modes
\begin{equation}
P_{\rm NLMG}\left(k^2\rightarrow -m_n^2\right)\approx \frac{+2}{\ell^2 (k^2+ m_n^2)}+\dots
\end{equation}
with masses
\begin{equation}\label{mn}
m_n=\frac{\pi}{2\ell} (2n-1)\, ,
\end{equation}
all of which have positive kinetic energy, following from the positivity of the residue at each pole \footnote{As argued in \cite{Aguilar-Gutierrez:2023kfn}, choosing the opposite sign in \req{nlmg1} produces a linearized spectrum which in turn  gives rise to an infinite tower of ghost-like modes. }. Again, this represents a generalization of the NMG case, for which one has a single massive mode of mass $m=1/\ell$.

{\bf Newtonian potential:} In the Newtonian approximation, the metric can be written as
\begin{equation}
\diff s^2=-\left(1+2\Phi\right)\diff t^2+\left(1-2\Psi\right)\diff \vec{x}^2\, ,
\end{equation}
and it is determined by two potentials $\Phi$ and $\Psi$, of which the former is the usual Newtonian potential. Assuming a pressureless stress energy tensor with energy density $\rho=T_{tt}$, these potentials satisfy the equations
\begin{align}
\Box\Psi+\frac{1}{2}F\left(\ell^2\Box\right)\ell^2\Box^2\left(\Phi+\Psi\right)&=-8\pi G \rho\, ,\\
\Box\left(\Psi-\Phi\right)&=-8\pi G \rho\, .
\end{align}
Notice that the second equation, coming from the trace of the equations of motion, is unchanged with respect to Einstein gravity (except for an overall sign). This is related to the fact that NLMG does not propagate scalar degrees of freedom. 

\begin{figure}[t!]
			\centering
			\includegraphics[width=0.48\textwidth]{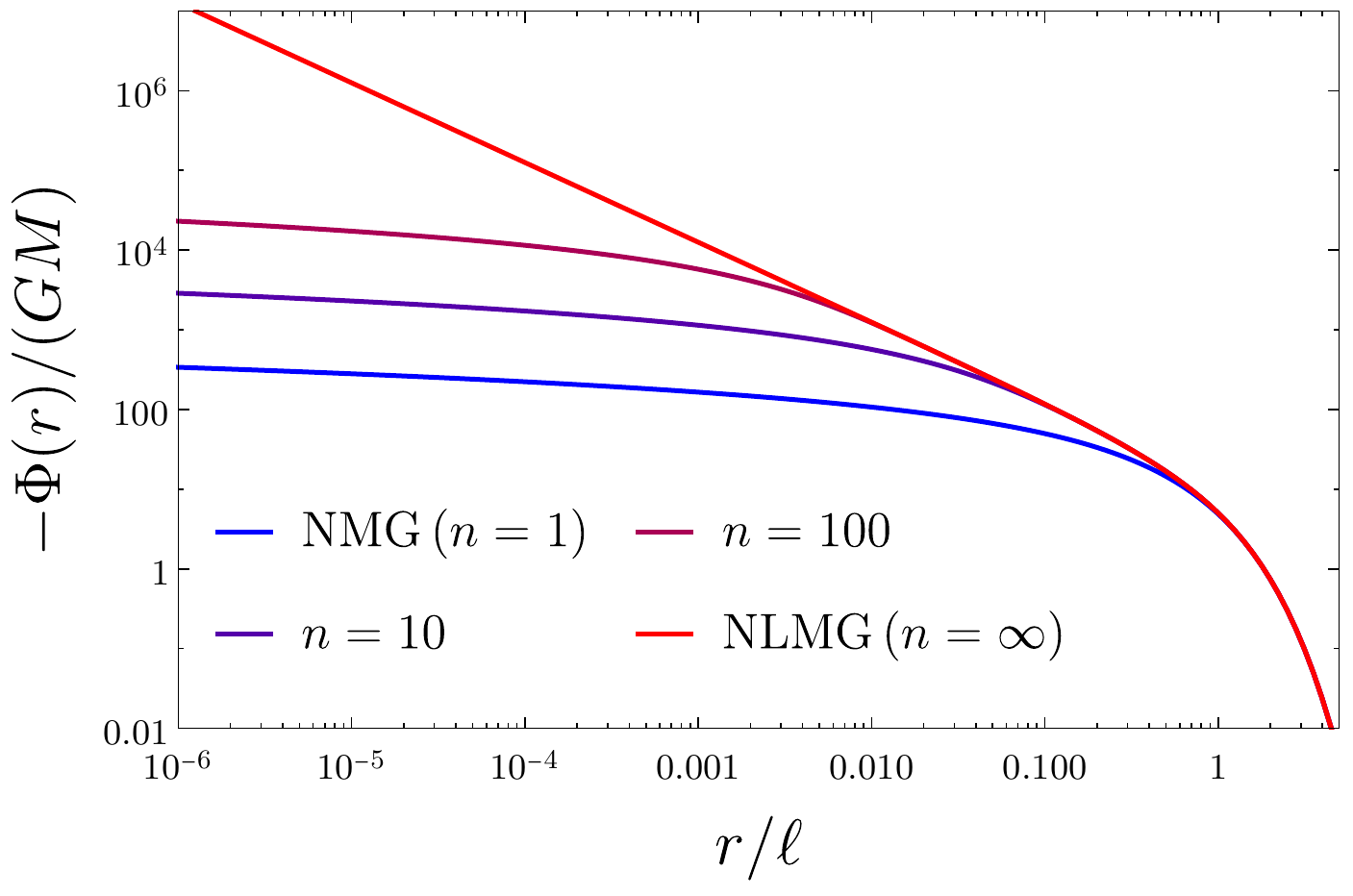}
			\caption{Newtonian potential for NLMG and for theories with $n=1$, $10$ and $100$ massive gravitons, obtained by truncating the sum \req{eq:Newtonianphi2}. In the log-log scale the NLMG potential appears as a straight line for $r\ll \ell$, indicating that $\Phi^{\rm NLMG}\propto 1/r$. For any finite number of gravitons the profile for $r\ll \ell$ is logarithmic instead. For large $r$ all the potentials decay exponentially.}
			\label{fig:Newtonpotential}			
\end{figure}

Let us then consider the case of a point particle, with $\rho=M \delta^{(2)}(\vec{x})$. The second equation implies that 
$\Psi-\Phi=-2GM \log(r/r_0)$, which can be seen to be (locally) equivalent to a gauge transformation.\footnote{However, this is a ``large'' gauge transformation, so it affects the global structure of the spacetime.} In the case of Einstein gravity, we furthermore have $\Phi=0$, so that the whole metric is flat. This is no longer the case in NLMG. By solving the equations in momentum space and then converting back to position space, one can see that the Newtonian potential is given by
\begin{equation}\label{eq:Newtonianphi}
\Phi^{\rm NLMG}(r)=-4\pi G\ell M\int_{0}^{\infty}\diff k\, J_{0}(k r) \tanh(\ell k)\, ,
\end{equation}
where $J_{0}(k r)$ is a Bessel function of the first kind. 

This formula secretly contains information on the massive degrees of freedom of the theory. To make this manifest, we utilize the following representation of the $\tanh$ function, 
\begin{equation}
\tanh(\ell k)=\frac{2}{\ell}\sum_{n=1}^{\infty}\frac{k}{k^2+m_{n}^2}\, ,
\end{equation}
where $m_n$ are precisely the masses in \req{mn}. Now, the integral in \req{eq:Newtonianphi} can be performed for each term in this series and we get 
\begin{equation}\label{eq:Newtonianphi2}
\Phi^{\rm NLMG}(r)=-8\pi G M\sum_{n=1}^{\infty}K_{0}(m_{n} r)\, ,
\end{equation}
where in this case, $K_{0}(mr)$ is a modified Bessel function of the second kind.  Clearly, this shows that the full Newtonian potential is a sum of the potentials from each individual massive graviton. In particular, the potential for NMG is given by a single term $\Phi^{\rm NMG}(r)=-8\pi G M K_{0}(mr)$.\footnote{Actually, there is an extra factor of two with respect to the result in NMG, but we will nevertheless refer to each term in \req{eq:Newtonianphi2} as the NMG potential.} 

The formula \req{eq:Newtonianphi2} is useful in order to determine the asymptotic behavior of the potential. Indeed, the Bessel function $K_0(m_n r)$ decays exponentially, and therefore for large distances the potential is dominated by the lightest mode. More precisely we have, 
\begin{equation}
\Phi^{\rm NLMG}(r\rightarrow\infty)= -\frac{8\pi^{3/2} G M}{\sqrt{2 m_{1} r}} e^{-m_{1} r}\, , 
\end{equation} 
which is the same behavior as in NMG with $m=m_1$. 

Let us then examine the behavior for small radius. For $mr\ll 1$, the modified Bessel functions posses a logarithmic divergence $K_0(m r)\sim -\log(m r)$. Thus, for NMG and for any theory with a finite number of massive gravitons the Newtonian potential diverges logarithmically near the origin. However, the result changes qualitatively when the full tower of massive modes is included. The easiest way to see this is to evaluate \req{eq:Newtonianphi} for $r\ll \ell$. In that regime, the main contribution to the integral comes from $\ell k\gg 1$ and hence the leading behavior is captured by setting $\tanh(\ell k) =1$.  Thus, we get
\begin{equation}
\Phi^{\rm NLMG}(r\ll \ell)=-\frac{4\pi G\ell M}{r}\, ,
\end{equation}
and remarkably, one recovers the usual $1/r$ behavior of a point particle in the four dimensional bulk space. 
As we noted, this cannot happen in any theory with a finite number of derivatives (hence a finite number of modes), so this is a genuine non-local effect. We illustrate this in Figure~\ref{fig:Newtonpotential} where we show how the Newtonian potential approaches that of NLMG as the number of massive gravitons is increased.  
We can also interpret this as the fact that the theory effectively becomes four dimensional at length scales shorter than $\ell$, where the massive modes become active.

{\bf Black holes?} The fact that the Newtonian potential exhibits a $1/r$ behaviour at short distances suggests that NLMG has asymptotically flat vacuum black holes. We have not been able to confirm this directly, but further support comes from recent developments in brane-world holography~\cite{Emparan:2020znc,Geng:2020qvw,Emparan:2022ijy}. 

In this context, the four-dimensional AdS $C$-metric has been used to describe three-dimensional, asymptotically AdS ``quantum BTZ'' black holes localized on a brane. The effective action on the brane consists of a gravitational theory coupled to a (cut-off) CFT. The gravitational sector of the theory is equivalent to NLMG with a cosmological constant and modified by an overall minus sign. 

\begin{figure}[t!]
			\centering
			\includegraphics[width=0.48\textwidth]{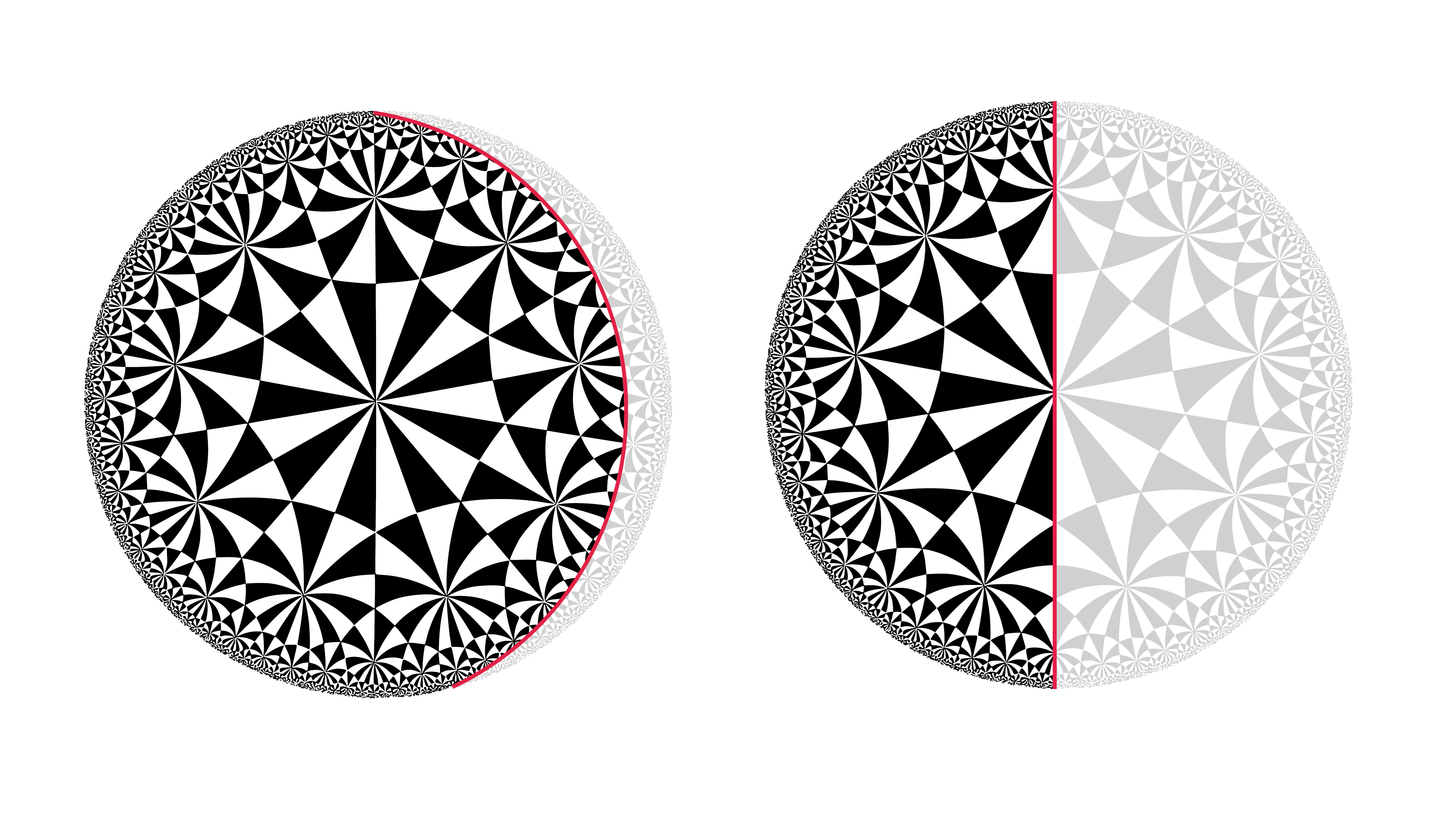}
			\caption{Visual depiction of brane-world holography. Left: A brane (in red) near the AdS boundary, representing the high-tension limit that approaches conventional AdS/CFT as tension drives the brane to the boundary. Right: The tensionless limit, where all CFT degrees of freedom are eliminated, leaving a purely gravitational, yet strongly coupled, brane theory.}
			\label{fig:quBTZ_brane}			
\end{figure}

The tension of the brane plays two roles. It governs the strength of the quantum back-reaction of the CFT on the brane geometry and also gives the cut-off of the CFT ---  see Figure~\ref{fig:quBTZ_brane}. In the limit of large tension, the gravitational dynamics on the brane turns off, the cut-off goes to zero, and one is left with a strongly coupled CFT on a non-dynamical BTZ black hole. On the other hand, in the tensionless limit, the CFT degrees of freedom are completely integrated out to yield a purely gravitational theory on the brane, albeit a strongly coupled one (\ie with large coupling constant). Crucially, this theory is not coupled to matter, so the overall minus sign relative to NLMG becomes irrelevant: solutions of the brane theory are also solutions of NLMG (with cosmological term) in this limit.

One can show that in the tensionless limit, the metric on the brane is an equatorial slice of the Schwarzschild-AdS$_4$ black hole  --- c.f. Section 2.3 of~\cite{Emparan:2020znc} --- with metric function $f(r) = 1 - 2m/r + r^2/\ell_4^2$. This confirms that the cosmological extension of NLMG does indeed contain black holes, at least in the limit of large coupling. Moreover, because of the $1/r$ term in the metric function, the existence of a horizon is not merely a consequence of the cosmological constant length scale. Since this term likely arises from the Newtonian potential, this all suggests that NLMG very likely contains black holes at finite coupling and in the asymptotically flat setting as well. 

{\bf Discussion:}
Non-Local Massive Gravity is a new unitary and non-local generalization of three-dimensional Einstein gravity and NMG motivated from top-down considerations. As we have seen, the linearized spectrum of the theory shows that it is perturbatively unitary at the quantum level, a fact which crucially relies on the trivial linearized dynamics of Einstein gravity combined with the unusual overall minus sign. Additionally, the presence of higher-derivative terms makes the theory power-counting UV-finite. 
However, analogously to NMG, we expect that the absence of scalar degrees of freedom makes the theory non-renormalizable \cite{Deser:2009hb, Muneyuki:2012ur}. In addition, we observe that the propagator of NLMG behaves as $P_{\rm NLMG}(k) \sim 1/k$ for $k\rightarrow\infty$  as opposed to the $1/k^4$ behavior of NMG or the $1/k^2$ of Einstein gravity. This makes NLMG even ``less renormalizable'' than those, which can ultimately be traced back to the fact that NLMG is somehow four-dimensional Einstein gravity in disguise.

The anomalous degree of divergence of the propagator of NLMG is also behind the unusal behavior of its Newtonian potential, which diverges as $1/r$, contrary to the logarithmic behavior of any truncation of the infinite tower of higher-derivative terms.
This, along with the aforementioned connections with the quantum BTZ solution and the four-dimensional ``origin'' of the theory suggest that NLMG admits non-trivial asymptotically flat black holes. In this regard, it is worth mentioning that non-local theories of gravity have been traditionally considered with the goal of resolving spacetime singularities \cite{Biswas:2005qr,Biswas:2011ar,Frolov:2015bta,Buoninfante:2018xiw, Boos:2018bxf, Boos:2020qgg}. The nature of our construction leads to a situation that turns out to be the opposite. This shows that non-local theories do not necessarily reduce the degree of divergence of spacetime singularities.

There are many possible future directions suggested by this work. Foremost among these would be studying the cosmological extension of NLMG, which would allow for direct holographic analysis of the theory. In our case, the construction of the quadratic action, which governs the linearized perturbations, relied crucially on the absence of a cosmological constant and so new techniques will be required to tackle this problem efficiently. It would also be of considerable interest to explicitly construct, or establish additional evidence for, black holes in the theory. Perhaps this could be achieved via a suitable generalization of the brane-world holography approach used in~\cite{Emparan:2020znc}, or it may be possible to utilize methods based on null shell collapse to understand the formation of mini black holes in the linearized regime~\cite{Frolov:2015bta}. Yet another possibility is to study the nature of curvature singularities in NLMG, which may provide a toy model for understanding the implications of quantum effects for the classical properties of singularities, such as ultra-locality and chaos~\cite{Belinsky:1970ew}.

\vspace{0.1cm}
\begin{acknowledgments} 
We thank Eric Bergshoeff and Roberto Emparan for useful comments. We also thank Sergio Aguilar-Gutierrez and Quim Llorens for collaboration on related topics. 
PB was supported by a Ram\'on y Cajal fellowship (RYC2020-028756-I) from Spain's Ministry of Science and Innovation. 
The work of PAC received the support of a fellowship from “la Caixa” Foundation (ID 100010434) with code LCF/BQ/PI23/11970032. 
The work of RAH received the support of a fellowship from ``la Caixa” Foundation (ID 100010434) and from the European Union’s Horizon 2020 research and innovation programme under the Marie Skłodowska-Curie grant agreement No 847648 under fellowship code LCF/BQ/PI21/11830027.

\end{acknowledgments}

\bibliographystyle{JHEP-2}
\bibliography{NLMG}
\noindent \centering
  
\end{document}
%